# The Ratio Between the Number of Sunspot and the Number of Sunspot Groups[1]

K. Georgieva[a], A. Kilçik[b], Yu. Nagovitsyn[c], and B. Kirov[a]
[a]SRTI-BAS, Sofia, Bulgaria
[b]Akdeniz University, Antalya, Turkey
[c]Central Astronomical Observatory at Pulkovo, St. Petersburg, Russia

**Abstract**—Data from three solar observatories (Learmonth, Holloman, and San Vito) are used to study the variations in the average number of sunspots per sunspot group. It is found that the different types of sunspot groups and the number of sunspots in these groups have different solar cycle and cycle to cycle variations. The varying ratio between the average number of sunspots and the number of sunspot groups is shown to be a real feature and not a result of changing observational instruments, observers' experience, calculation schemes, etc., and is a result of variations in the solar magnetic fields. Therefore, the attempts to minimize the discrepancies between the sunspot number and sunspot group series are not justified, and lead to the loss of important information about the variability of the solar dynamo.

## 1. INTRODUCTION

Sun is the main source of energy for the Earth's system, providing orders of magnitude more energy than all other extraterrestrial sources taken together (irradiance from distant stars, gamma ray bursts, galactic cosmic rays, etc.). Sun emits a continuous but variable flow of matter (the solar wind – the ever expanding solar corona), recurrent high speed solar wind streams from long-living solar coronal holes which persist for several solar rotations and bathe the Earth each time their source region is in a geoeffective position, outbursts of plasma with embedded magnetic fields from the corona (coronal mass ejections), solar energetic particles associated with strong impulsive solar events, and electromagnetic radiation across virtually the entire range of wavelengths (the total solar irradiance), with sporadic sudden, rapid, and intense variations in brightness (solar flares). All variations in the energy output and the associated outlook of the Sun are defined as "solar activity". They are all manifestations of the solar magnetic field which is in turn the result of the action of the solar dynamo.

There are different indicators of solar activity, reflecting different solar processes and using different direct or indirect measurable or proxy parameters. The number of sunspots and the number of sunspot groups are the longest instrumental data records of solar activity. Sunspots – dark spots on the solar surface – do not themselves in any way affect the Earth, but they reflect the magnetic activity of the Sun and are related to geoeffective solar events, so they have been widely used to evaluate the long-term evolution of solar activity, and its effects on the terrestrial system.

The original "relative sunspot number", known also as "Wolf number" or "Zurich international sunspot number", $R_Z$, was defined by Wolf as

$$R_Z = k(10g + n), \qquad (1)$$

where $n$ is the number of individual sunspots, $g$ – the number of sunspot groups, and $k$ is the correction factor for each observer accounting for the differences in instruments, measurement techniques, viewing conditions, observers' experience, etc. (Waldmeier, 1961). The yearly/monthly values of the $R_Z$ series cover the period from 1700/1749, respectively, to June 2015.

The group sunspot number $R_G$ was introduced by Hoyt and Schatten (1998). It is based on the parameter $g$ in equation (1) – the number of sunspot groups which is more reliably determined and allows the inclusion of earlier observations, so it expands the record back to the earliest telescopic observations in 1610. $R_G$ is defined as

$$R_G = \frac{12.08}{N}\sum k_i' G_i, \qquad (2)$$

where $G_i$ is the number of sunspot groups observed by the $i$-th observer, $k_i'$ is the $i$-th observer's correction factor, $N$ is the number of observations used to calculate $R_G$, and 12.08 is a normalization number chosen to make the mean $R_G$'s identical with the mean $R_Z$'s for the period 1874 to 1976 when the Royal Greenwich Observatory actively made sunspot observations (Hoyt and Schatten, 1998).

[1] The article is published in the original.

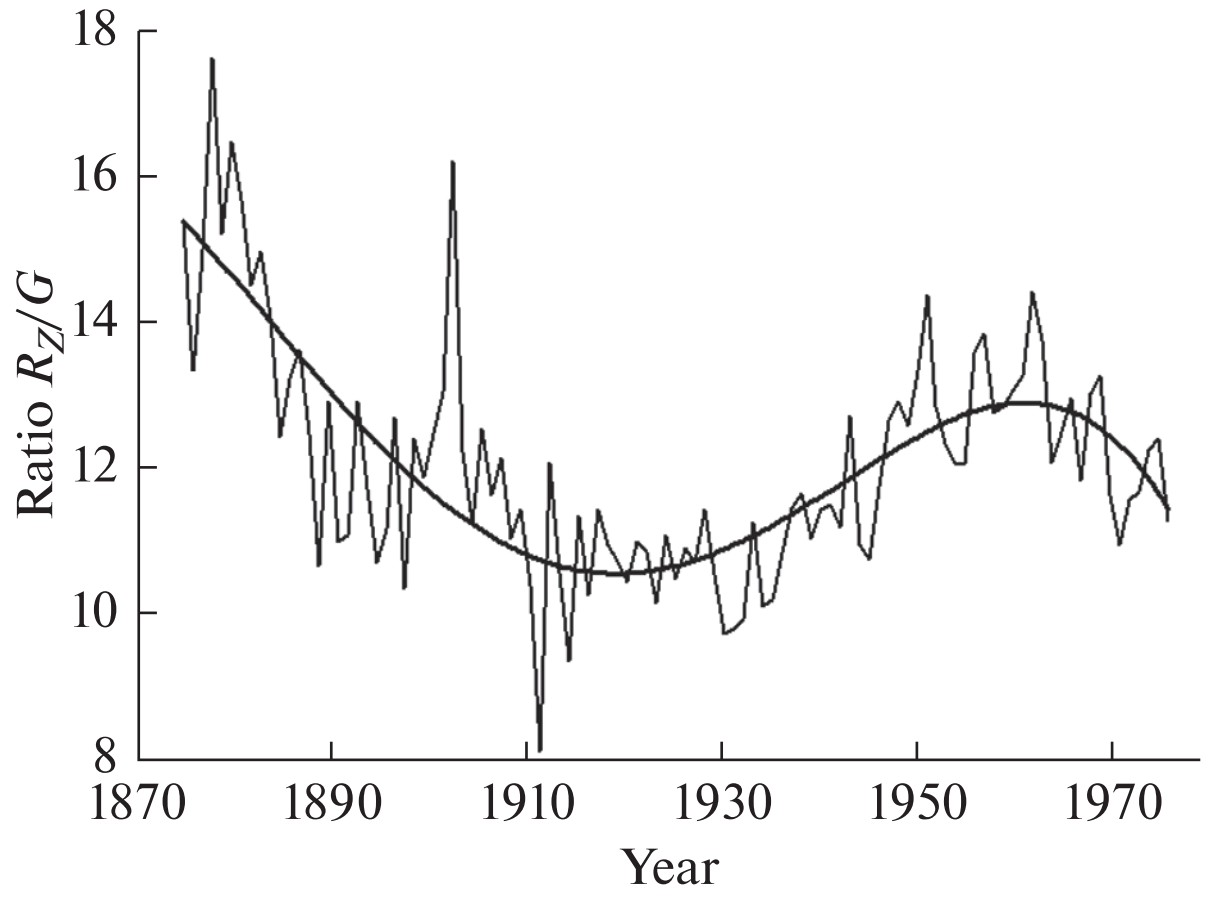

**Fig. 1.** Variations in the ratio between the International Zurich sunspot number $R_Z$ and the number of sunspot groups.

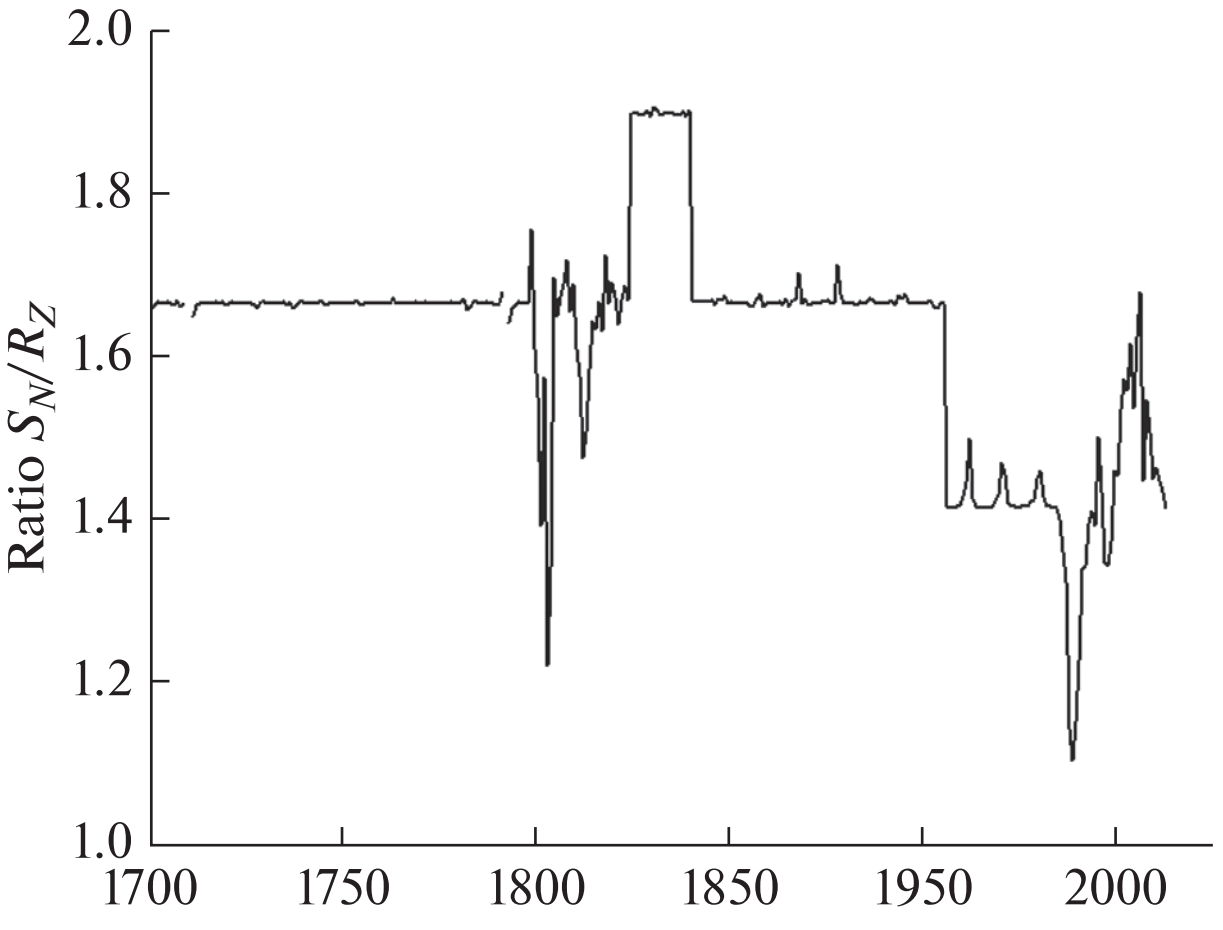

**Fig. 2.** The corrections applied to $R_Z$ to transform it into the new series $S_N$.

However, even in this limited period, the ratio between $R_Z$ and $R_G$ is not constant, and displays long-term quasi-cyclic variations (Fig. 1).

On longer time-scales, both $R_Z$ and $R_G$ have upward trends since the Maunder minimum, but the trend is larger for $R_G$ than for $R_Z$. This means that using $R_G$ would imply larger long-term solar variability, and consequently more important impact of the Sun on the terrestrial variability, for example climate changes, than using $R_Z$.

The different trends of the two sunspot indices have inspired efforts to recalibrate the two time series with the aim to "rectify discrepancy between Group and International sunspot number series", and to publish "a vetted and agreed upon single sunspot number time series" (Cliver et al., 2013; Clette et al., 2014), because "given the importance of the reconstructed time series, the co-existance of two conflicting series is a highly unsatisfactory solution which should be actively addressed" (Cliver et al., 2013).

In September 2011, a "Sunspot Number (SSN) Workshop" was held in Sunspot, New Mexico, USA to address this question, followed by several other similar workshops during 2012–2014 (http://ssnworkshop.wikia.com/wiki/Home). The net result of this activity was that since July 1st 2015, the Sunspot Index Data Center in Brussels terminated the more than 400 year long data series of the International relative sunspot number $R_Z$, and replaced it by "a new entirely revised data series" $S_N$ (http://www.sidc.be/silso/datafiles). The Sunspot Index Data Center was created in 1980 in the Royal Observatory of Belgium as a World Data Center with the task to continue the International relative sunspot number record after the decision of the new director of the Swiss Federal Institute of Technology J.O. Stenflo to terminate the 130-year-long Zürich sunspot number observational program initiated by R. Wolf (Clette et al., 2007).

Figure 2 illustrates the corrections applied to $R_Z$ to transform it into the new series $S_N$. As a result of this transformation, the overall level of solar activity as measured by $S_N$ was significantly increased as compared to $R_Z$ (the ratio $S_N/R_Z$ is always greater than 1), and the "Modern Grand Maximum" in the second half of the 20th century (Usoskin et al., 2007) was reduced to an ordinary secular maximum of solar activity, not significantly different from the secular maxima in previous centuries (the ratio since 1950 is much lower than the ratio from 1700 to 1950). Consequently, the Sunspot Group Number data series $R_G$ was also reconstructed, and a new series $G_N$ was produced (Svalgaard and Schatten, 2016), which matches the new Sunspot Number series $S_N$ very closely, and both indices have practically no long-term trends, unlike the original $R_Z$ and $R_G$. This has important implications for the evaluation of solar activity effects on terrestrial processes. For example, at a press briefing during the IAU XXIX General Assembly in 2015 it was announced that the corrected sunspot history suggests that "rising global temperatures since the industrial revolution cannot be attributed to increased solar activity" (https://www.iau.org/news/pressreleases/detail/iau1508/).

In this way, the goal to "rectify discrepancy between Group and International sunspot number series" was fulfilled. But the other stated goal – to publish an "agreed upon single sunspot number time series" was not achieved, neither for the sunspot number nor for the sunspot group number. On the contrary, the presentation of the two new series was met by vigorous criticism, and led to the ongoing creation of more and more still newer alternative series by scien-tists not only outside the Sunspot Workshops initiative group, but even by members of this very group [http://www.spaceclimate.fi/SC6/presentations/session2b/Frederic_Clette_SC6.pdf].

Clette et al. (2014) summarized the possible flaws in the $R_Z$ and $R_G$ series, and justified some of the corrections made to transform $R_Z$ into $S_N$, and Svalgaard and Schatten (2016) presented the "backbone method" used to construct $G_N$. In the present study we are not dealing with the sources of uncertainties in the original $R_Z$ and $R_G$ series, neither with the applied corrections to remove the possible flaws, nor with the reliability of $S_N$, $G_N$, and the other newly published series. Instead, we are concentrating on the discrepancies between the long-term variations of the original $R_Z$ and $R_G$ series with the goal to estimate whether the lack of discrepancies in the newly constructed $S_N$ and $G_N$ series is a real feature, or an artifact of the goal to "reconcile" them.

In Section 2 we describe the data we use, and their processing. We present our results in Section 3, and summarize and discuss them in Section 4.

## 2. DATA AND METHODS

Using data for the four most recent solar cycles, Kilcik et al. (2014) separated active regions into four types, based on the size of the sunspot group and the sunspot evolution:

S – simple groups (Zurich classes A and B), in the early stage of their evolution with tiny spots which do not have penumbrae.

M – medium (class C); in the middle of their group evolution with two or more spots which demonstrate bipolarity and have a penumbra at one end of the group.

L – large (classes D, E, and F); well developed groups spreading from 10 to over 15 degrees of solar longitude, with two or more bipolar spots, with penumbrae at both sides of the group.

F – final (class H) types; the decayed remnants of M and L groups, containing a single spot group with penumbra occasionally accompanied by a few small spots.

Obviously, a varying relative abundance of these four types of sunspot groups would imply a varying ratio between $R_Z$ and $R_G$ calculated according to equations (1) and (2), because of the term $n$ in equation (1) – the number of sunspots in the group, which is different in the different types of sunspot groups.

Possible sources of uncertainties in the long time series of both the International sunspot number and the Group sunspot number are the inevitable changes of observers during the more than four centuries over which the measurements were collected, the changes of pilot observatories, instruments, observational routines, calculation schemes, etc. Therefore, using data from observatories with continuous and homogeneous (though much shorter) data records could give a clue whether the possible variations in the ratio between the International Sunspot Number and the Group Sunspot Number are a real feature or a result of the above reasons.

The data we use are from the National Geophysical Data Center (ftp://ftp.ngdc.noaa.gov/STP/SOLAR_DATA) which provides various active region parameters including the sunspot group (SG) classification and sunspot counts – the number of sunspots on the solar disk (SSC) determined for each day. The database collected by the US Air Force/Mount Wilson Observatory includes measurements from Learmonth (LEAR), Holloman (HOLL), and San Vito (SVTO) Solar Observatories. In a recent study [Georgieva et al., 2016] we used the LEAR data as the principal data source, while the data gaps were filled in with observational records from one of the other stations, so that a nearly continuous time series was produced. Here, we study separately the data from all three observatories.

The data were processed in the following way: First, we separated sunspot groups into the four categories S, M, L, and F according the classification of Kilcik et al. (2014), and calculated the daily total number of sunspot groups (SSGs) and the sunspot counts (SSCs) for each category. Then we calculated monthly averages from the daily numbers for those categories, thus obtaining monthly values, essentially independent of data gaps. We summed all categories to find total monthly numbers of SSCs and SSGs. To remove short-term fluctuations and to reveal long-term trends, the monthly averaged time series were smoothed with 12-month moving average. Further, we calculated the fraction of S, M, L, and F groups in the total number of SSGs, and the fractions of sunspots in each group in the total number of SSCs.

The data from LEAR and HOLL cover the time interval from January 1982 to December 2015: the descending branch of cycle 21 (1982–1986), cycles 22 (1986–1996) and 23 (1996–2008), and the first half of cycle 24. SVTO data start in January 1986 and also span to December 2015, covering cycles 22 (1986–1996) and 23 (1996–2008), and the first half of cycle 24.

## 3. RESULTS

Figure 3 presents the average number of sunspots per sunspot group for LEAR, HOLL, and SVTO. For comparison, the total number of sunspots averaged over the three observatories is added. There are some differences in the values calculated from the records of the different observatories, especially around sunspot maximum, but the general pattern is the same in all three observatories: The ratio SSC/SSG has strong solar cycle variations, and varies from cycle to cycle. Except for SVTO, it is almost equal in the maxima of cycles 22 and 23. In all three observatories the ratio is significantly smaller in the maximum of cycle 24 than in the previous two maxima.

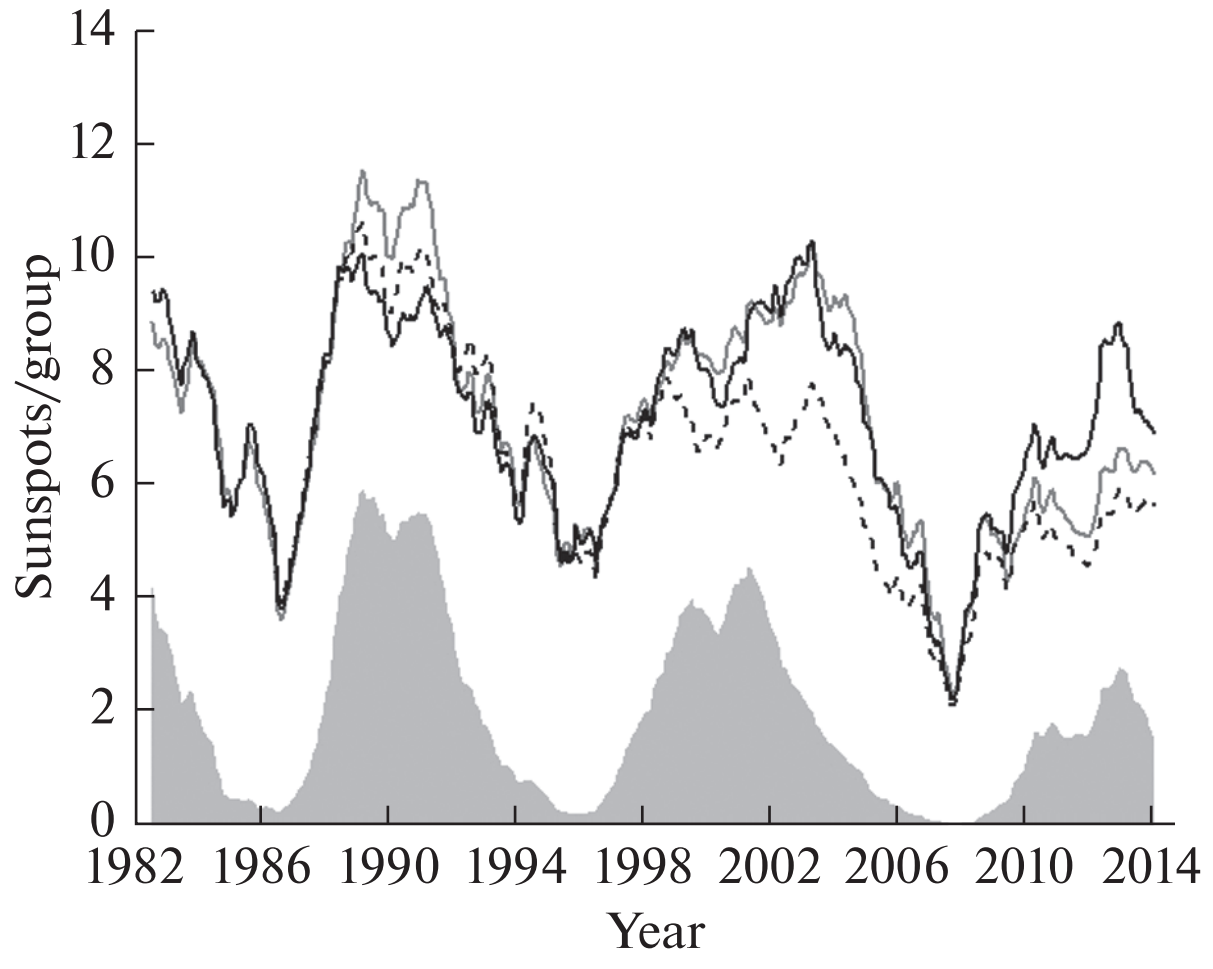

**Fig. 3.** Total number of sunspots per sunspot group measured by LEAR (black solid line), HOLL (grey solid line), and SVTO (black dotted line), 12-point smoothed monthly values, compared to the total number of sunspots averaged over the three observatories (grey shading), 12-month smoothed monthly values.

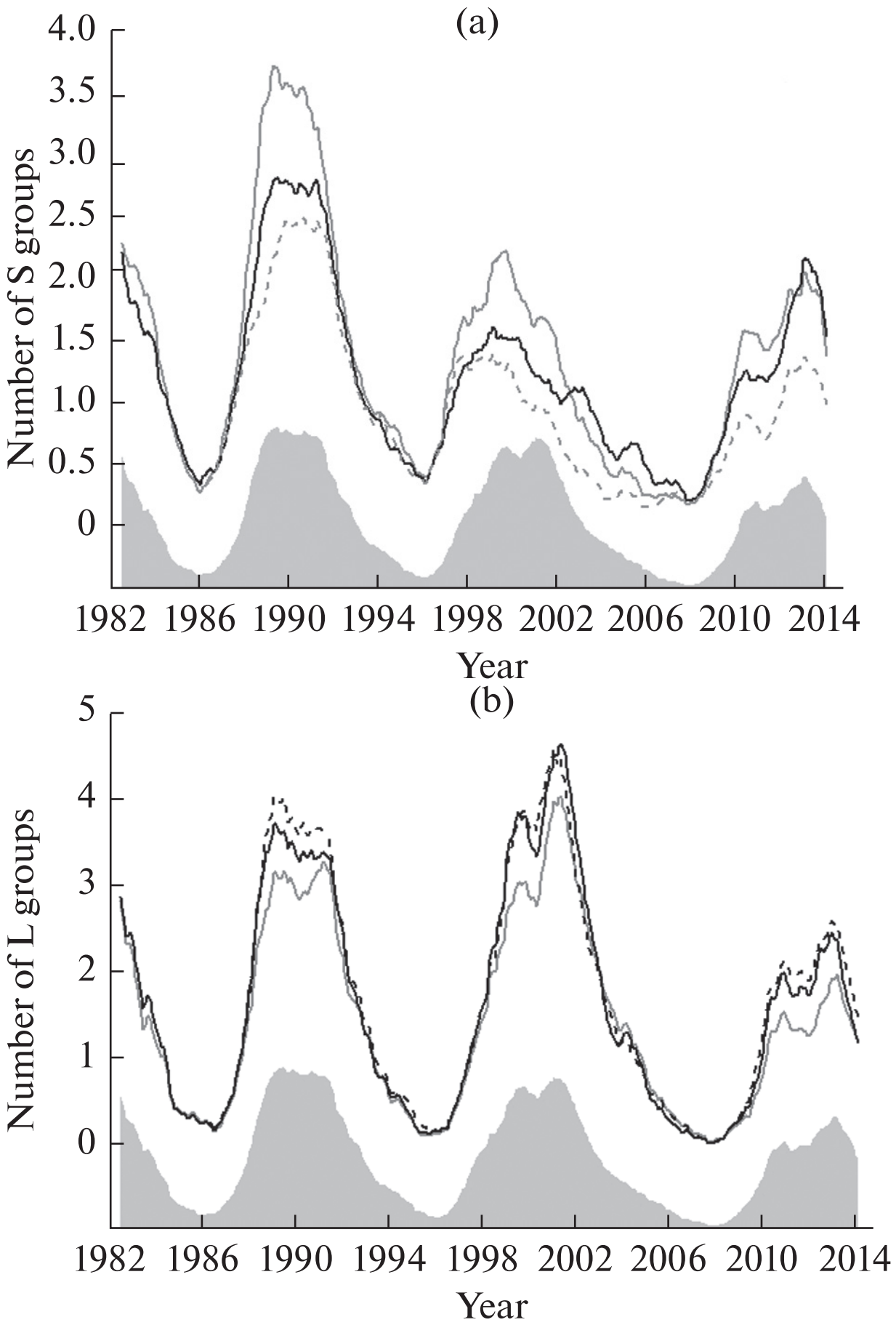

**Fig. 4.** Number of (a) S and (b) L sunspot groups measured in LEAR (black solid line), HOLL (grey solid line), and SVTO (black dotted line), and the total number of registered groups, averaged over the three observatories (grey shading), 12-month smoothed monthly averages.

During the course of the solar cycle, as well as from cycle to cycle, the relative number of the groups in the different categories varies. Figure 4 illustrates the number of S and L groups measured in the three observatories, with the total number of sunspot groups averaged over the observatories added for reference. (Here and further we concentrate on these two categories which are the two extremes and are the most indicative of the changes in the operation of the solar dynamo). It is a well known fact that the total number of sunspot groups peaks in cycle maxima and drops in cycle minima, and this is also true for the number of groups in the each category. From Fig. 4 it can be seen that the contribution of the different groups to the total number of groups varies from cycle to cycle. The number of S groups in the maxima of cycles 23 and 24 is much lower than in the maximum of cycle 22, and probably lower than in the maximum of cycle 21 (which is not complete in our data records). There is no substantial difference between the maxima of cycles 23 and 24. The number of L groups is highest in the maximum of cycle 23 and almost two times lower in the maximum of cycle 24.

This is further demonstrated in Fig. 5 where the fraction of S and L groups is shown. In all observatories, the fraction of registered L groups follows the sunspot cycle, with broad maxima around sunspot maxima, and sharp minima in sunspot minima. The variations of the fraction of S groups are opposite: broad minima in sunspot maxima, and sharp maxima in sunspot minima. The total number of sunspot groups is strongly dominated by the small groups in solar minimum, and the fraction of small groups has been increasing in the last three minima, being almost two times higher in the minimum between cycle 23 and 24 than in the minimum between cycles 21 and 22. This high relative abundance of S groups groups in the minimum between cycle 23 and 24 is accompanied by a sharp decrease of the abundance of L groups, and as a result the total number of sunspot groups is almost entirely determined by the number of small groups.

The fraction of L groups has decreased in the maximum of cycle 24 as compared to the maximum of cycle 23, and the fraction of S groups has increased. As a result, their contribution is almost equal in cycle 24 maximum, while the fraction of L groups was more than twice higher than that of S group in cycle 23 maximum. The fractions of M and F groups remained about the same in these two solar maximum periods (not shown). A continuous decrease in the fraction of L groups and an increase in the fraction of S groups are observed in consecutive solar minima. In the deepest minimum between cycles 23 and 24 the contribution of small sunspot groups to the total number of groups was an order of magnitude higher than that of large groups, as compared to the factor of 3 to 5 for the different observatories in the previous two minima.

Not only the relative contribution of different sunspot groups to the total SSG number, but also the fraction of sunspots contained in different groups vary in the course of the sunspot cycle and from cycle to cycle (Fig. 6). Around sunspot maximum, the total number of sunspots is strongly dominated (~80%) by the sunspots in large groups with ~10% from sunspots in small groups, while in sunspot minima the contribution of sunspots in small groups increases. During the previous two minima, the fraction of sunspots in S groups was roughly equal to the fraction of sunspots in L groups. In the minimum between cycles 23 and 24 the contribution of sunspot in small groups to the total number of sunspots was a factor of 7 to 8 higher than that of sunspots in large groups.

## 4. SUMMARY AND DISCUSSION

Kilcik et al. (2011, 2014), based on data from Learmonth Observatory, showed that different types of sunspots and groups of sunspots behave differently over a solar cycle, and their relative abundance varies from cycle to cycle. Here we confirm and expand this result for two more observatories: Holloman and San Vito. We also find that:

(1) Different types of sunspot groups have different solar cycle and long-term variations. Large groups prevail around sunspot maximum, and the small groups prevail in sunspot minimum;

(2) In a broad interval around solar maximum, the total sunspot number is strongly dominated by sunspots contained in large groups. In a short period around solar minimum, the contribution of sunspots in small groups increases and becomes comparable to or even higher than that of sunspots in large groups.

(3) In the minimum between cycles 23 and 24, the relative prevalence of small groups and of sunspots in small groups were both much higher than in the previous two minima.

(4) For both small and large sunspot groups, the average number of sunspots in a sunspot group varies in the course of the sunspot cycle, being maximum in sunspot maximum and minimum in sunspot minimum.

(5) The ratio between sunspot counts and sunspot groups varies also from cycle to cycle. In the last three solar maxima the number of sunspots per group has been continuously decreasing, and is much smaller in cycle 24 than in the previous two cycles.

Our results are in agreement with Tlatov (2013) who, based on data from Kislovodsk observatory, showed that the variation in the average number of sunspots in one group has a trend, and this number decreased from cycle 19 cycle 24, moreover this ratio has cyclic variations with a period of about 100 years. This is also what is seen in the original $R_Z$ and $R_G$ series (Fig. 1).

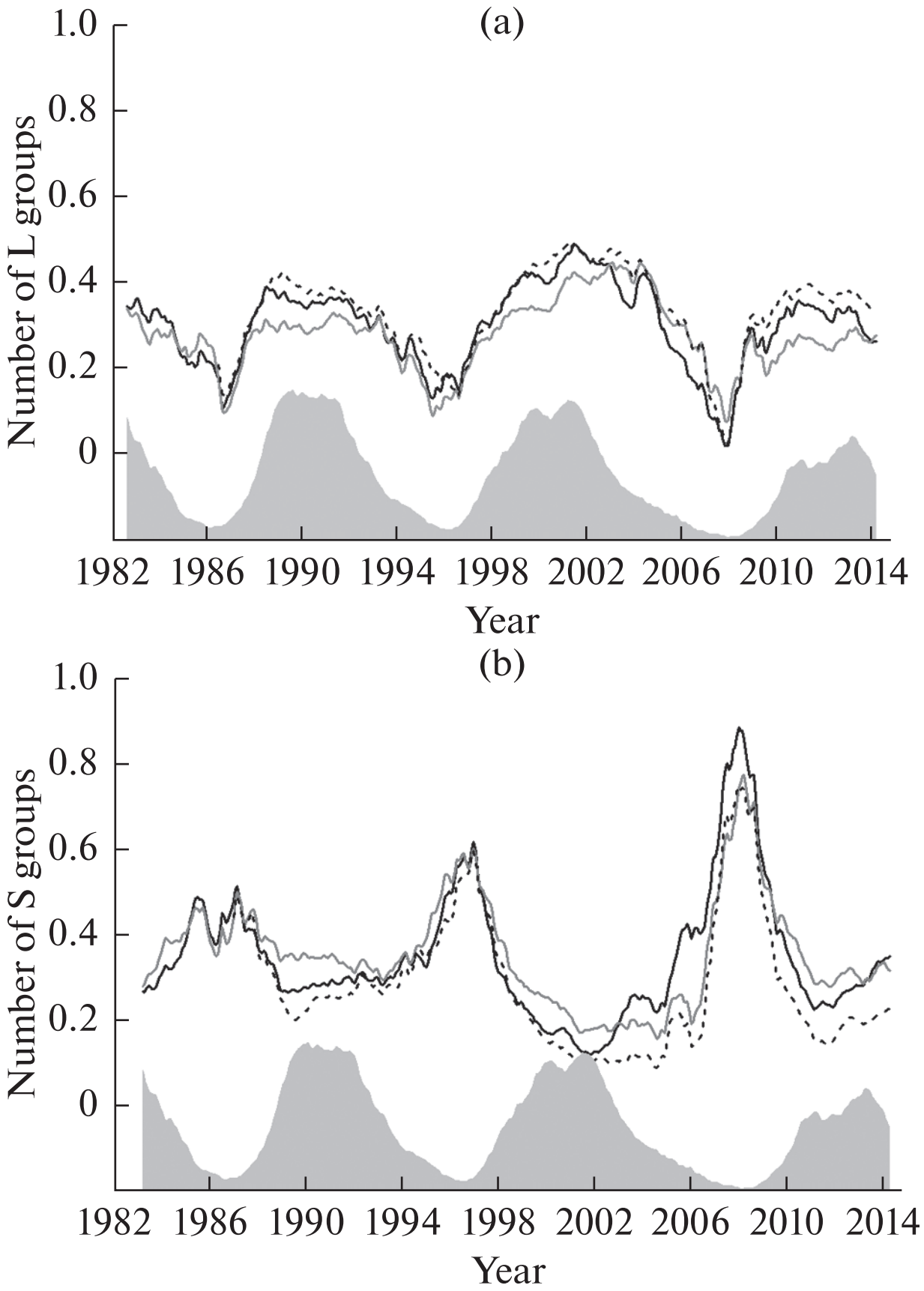

**Fig. 5.** Variations of the fraction of S groups (grey) and in L groups (black) in LEAR (thick solid line), HOLL (thick dotted line), and SVTO (thin solid line), with the total number of registered groups, averaged over the three observatories (grey shading) added for comparison, 12-month smoothed monthly averages.

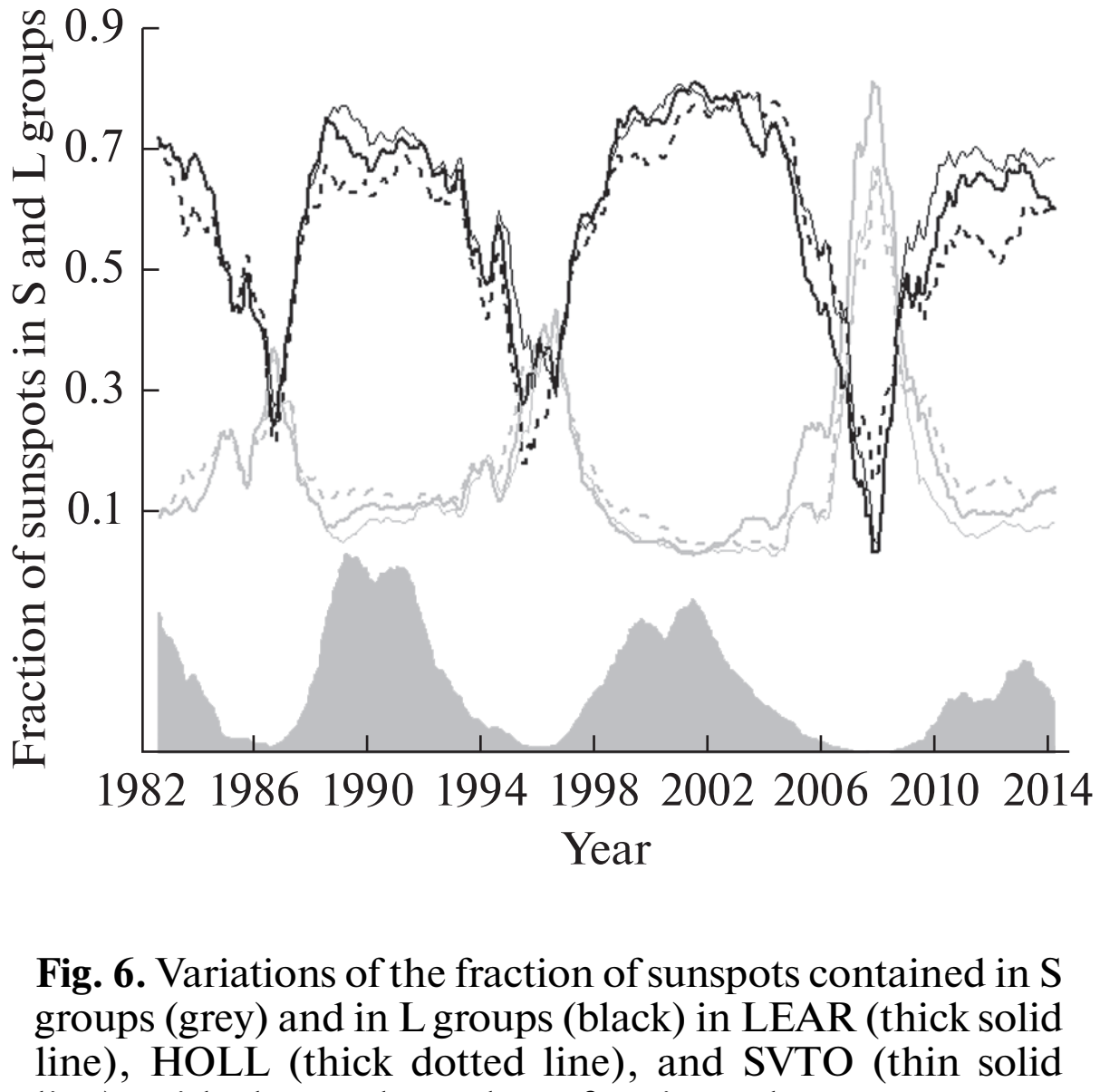

**Fig. 6.** Variations of the fraction of sunspots contained in S groups (grey) and in L groups (black) in LEAR (thick solid line), HOLL (thick dotted line), and SVTO (thin solid line), with the total number of registered sunspots, averaged over the three observatories (grey shading), 12-month smoothed monthly averages.

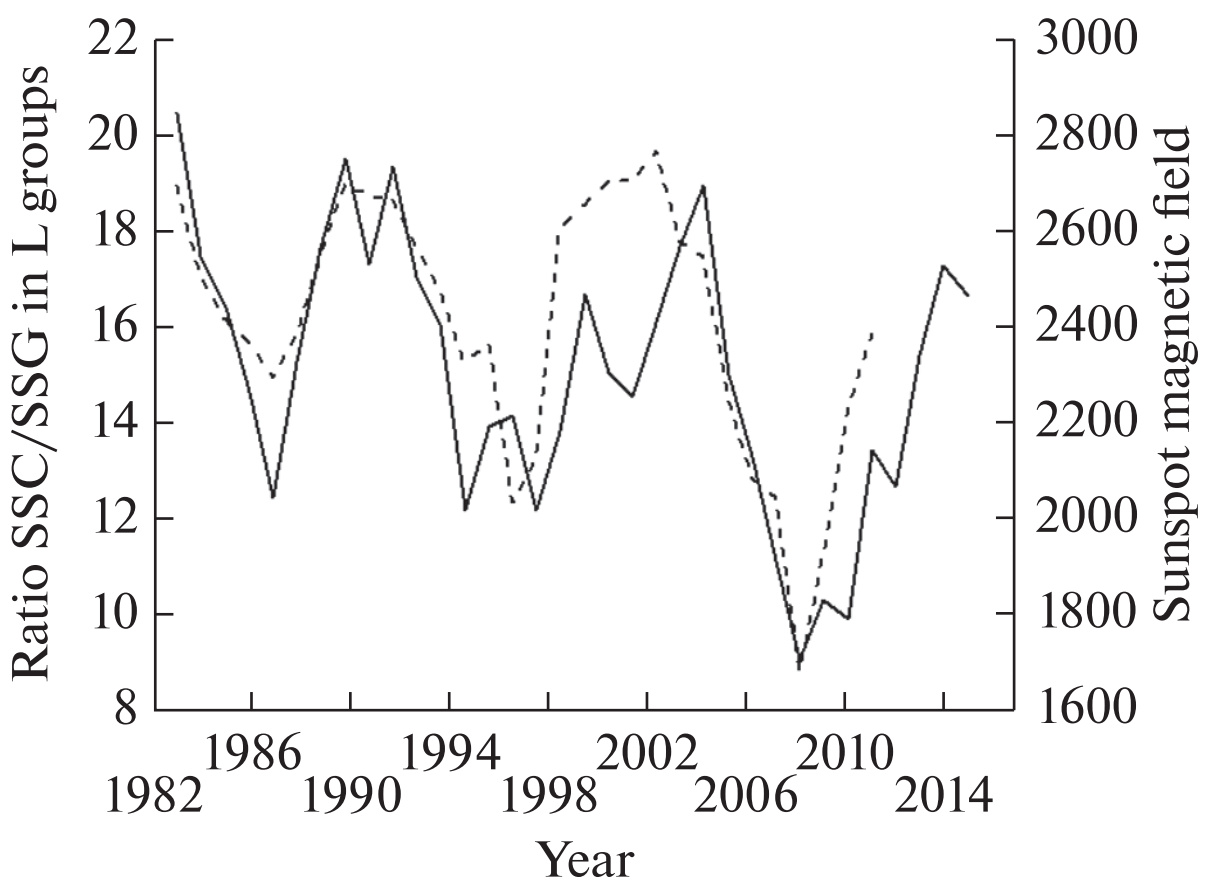

**Fig. 7.** Variations of the average number of sunspots in large sunspot groups (solid line) and of the sunspot magnetic fields (dashed line), yearly averages.

On the other hand, Curto et al. (2016) used the historical heliophysical series collected at the Ebro Observatory in Spain to study the ratio between sunspots and sunspot groups, and found that this ratio was quite stable, with neither sunspot cycle, nor longer term variations. This stability is in agreement with the recently presented $S_N$ and $G_N$ series. But, to the best of our knowledge, this is the only result based on data from an individual observatory, which shows no cyclic variations and no trend in the ratio between sunspot counts and sunspot groups. The discrepancy between Erbo, on the one hand, and Kislovodsk, Learmonth, Holloman and San Vito, on the other hand, deserves detailed investigation.

What can be the physical reasons for the varying ratio between $R_Z$ and $R_G$?

Sheeley (1966), using the then newly developed by Leighton method of photographically mapping of solar magnetic fields (Leighton, 1959) found that the complexity (respectively, the number of sunspots) of a sunspot group increases with its magnetic field. Pevtsov et al. (2012) based on historic synoptic data sets from seven observatories in the former USSR, found that the sunspot field strengths vary cyclically with maxima around sunspot maxima and minima around sunspot minima, with no indication of a secular trend in the maxima of sunspot cycles 19–23. On the other hand, Penn and Livingston (2011), using the NSO Kitt Peak McMath-Pierce telescope, found that the magnetic field in sunspots has been decreasing in time since 1990's, but with no dependence on the solar cycle. The explanation of this contradiction was offered by Nagovitsyn et al. (2012): While Pevtsov et al. (2012) used only the biggest sunspots for the analysis, Penn and Livinston (2011) used all visible sunspots. During the period of 1998–2011, the number of large sunspots whose magnetic fields do show sunspot cycle variations but no long-term trend, gradually decreased, while the number of small sunspots which have weaker magnetic fields decreasing in time but without sunspot cycle dependence, steadily increased.

This can be also the explanation of the solar cycle variations of the ratio between $R_Z$ and $R_G$. Big sunspots whose magnetic fields display solar cycle variations, are contained in large sunspot groups. Cyclic variations in the magnetic field in large sunspot groups would lead to cyclic variations not only in the magnetic fields of the sunspots in these groups, but also in the number of sunspots in these large groups, as found by Sheeley (1966). To check this, we have compared the number of sunspots per sunspot group in large groups to the big sunspots' magnetic fields as measured by the network of observatories in the former USSR (Fig. 7). The correlation between the ratio SSC/SSG and the sunspots' magnetic field is 0.79 with $p < 0.001$. For comparison, this correlation for the small groups, though statistically significant, is only 0.48 (not shown), in agreement with the result of Nagovitsyn et al. (2012).

And, as the total number of sunspots during sunspot maximum and in a broad interval around it (Fig. 6), is strongly dominated by the sunspots in large groups, the ratio between the average number of sunspots per sunspot group is dominated by the ratio SSC/SSG in large groups.

Long-term variations in the ratio between $R_Z$ and $R_G$ may be due to the variations in the relative abundance of large and small sunspot groups. We can expect that small sunspot groups' magnetic fields have long-term variations matching the long-term variations in small sunspots' magnetic fields, and the varying magnetic fields would lead to varying number of sunspots in a group. With increasing portion of small sunspot groups and decreasing number of large sunspot groups, the influence of the ratio SSC/SSG in small groups on the average number of sunspots per sunspot group will increase, and the ratio $R_Z/R_G$ will decrease. The opposite will be true when the portion of large sunspot groups increases.

In conclusion, our results unambiguously demonstrate that the variable ratio between the number of sunspots and the number of sunspot groups is a real feature, and a manifestation of the variations in the operation of the solar dynamo. Therefore, the attempts to minimize the differences between these two data series are not justified, and the resulting new closely matching "recalibrated" series are most probably erroneous. Instead of trying to "rectify the discrepancies", the variations in the ratio between the number of sunspots and the number of sunspot groups should be investigated with the aim to derive additional information about the long-term evolution of the Sun and the solar dynamo.

SPELL: 1. OK